\documentclass[preprint,12pt]{elsarticle}

\usepackage{amssymb}

\usepackage{booktabs}
\usepackage{graphicx}
\usepackage{algorithm}
\usepackage{algpseudocode}
\usepackage{setspace}

\journal{Future Generation Computer Systems}

\begin{document}

\begin{frontmatter}



\title{Parallel Processing of Large Graphs}


\author{Tomasz Kajdanowicz, Przemyslaw Kazienko, Wojciech Indyk}

\address{Wroclaw University of Technology, Poland}

\begin{abstract}
More and more large data collections are gathered worldwide in various IT systems. Many of them possess the networked nature and need to be processed and analysed as graph structures. Due to their size they require very often usage of parallel paradigm for efficient computation. Three parallel techniques have been compared in the paper: MapReduce, its map-side join extension and Bulk Synchronous Parallel (BSP). They are implemented for two different graph problems: calculation of single source shortest paths (SSSP) and collective classification of graph nodes by means of relational influence propagation (RIP). The methods and algorithms are applied to several network datasets differing in size and structural profile, originating from three domains: telecommunication, multimedia and microblog. The results revealed that iterative graph processing with the BSP implementation always and significantly, even up to 10 times  outperforms MapReduce, especially for algorithms with many iterations and sparse communication. Also MapReduce extension based on map-side join usually noticeably presents better efficiency, although not as much as BSP. Nevertheless, MapReduce still remains the good alternative for enormous networks, whose data structures do not fit in local memories.
\end{abstract}

\end{frontmatter}


\section{Introduction}
Many technical and scientific problems are related to data with the networked nature, which can be relatively simply represented by means of graph structures. Graphs provide a very flexible abstraction for describing relationships between discrete objects. Many practical problems in scientific computing, data analysis and other areas can be modelled in their essential form by graphs and solved with the appropriate graph algorithms. 

In many environments graph structures are so big that they require specialized processing methods, especially parallel ones. This becomes particularly vital for data collections provided by users leaving their traces in various online or communication services like multimedia publishing portals or social networking sites, e.g. YouTube or Facebook. Additionally, these datasets reflect various user behaviour, so their graph representation may be complex with multiple relationships linking network nodes. This requires analytical methods dealing not only with simple graphs but also hypergraphs or multigraphs.

As graph problems grow larger in scale and more ambitious in their complexity, they easily outgrow the computation and memory capacities of single processors. Given the success of parallel computing in many areas of scientific computing, parallel processing appears to be necessary to overcome the resource limitations of single processors in graph computations. Parallel graph computation is, however, challenging \cite{challenges} and before the advent of cloud computing and Hadoop, programmers had to use ill-suited distributed systems or design their own systems, which required additional effort to provide fault-tolerance and to address other problems related with parallel processing \cite{pregel}. The rise of the MapReduce concept and Hadoop - its open source implementation provided the  researchers a powerful tool to process large data collections. Recently, Hadoop has become a de facto standard in academia and a significant solution for parallel processing in industry. It has been used in various areas, including some graph processing problems \cite{yang:efficient}. 

The MapReduce model is, however, badly suited for iterative and graph algorithms. There has been a lot of research in creating design patterns improving MapReduce performance for graphs like \cite{lin}, \cite{dyer}, or building systems that would aid iterative processing on MapReduce \cite{haloop}, \cite{pegasus}, \cite{imapreduce}, \cite{surfer}, \cite{ihadoop}. Google reacted to that by development of Pregel \cite{pregel} –-- an alternative system that implements the Bulk Synchronous Parallel (BSP) programming model \cite{valiant} for graph processing. 

The main difference in processing of regular data structures (tables) and relational models (graphs) relies in different problem decomposition. Processing table structures is composed of handling of individual records (rows in the table). For the networked data, single processing of a graph vertex usually requires the access to the neighbourhood of this vertex, which for most algorithms remains fixed for the whole processing time. This data may be either accessed at every algorithm iteration via distributed file system (e.g. HDFS) as in the case of MapReduce or preserved locally for the entire processing - the case of BSP.

Both different parallel processing methods, i.e. MapReduce and BSP along with the map-side join MapReduce modification have been implemented in the Hadoop environment --- they three were used in the experiments presented in this paper. Each approach was independently applied to solve two distinct graph analytical problems: single source shortest path (SSSP) calculation and collective classification of network vertices with relational influence propagation (RIP). The graph algorithms had iterative nature what enabled testing their various parallel implementations in the following steps. The iterative computation was carried out in cloud environments containing various number of machines to compare scalability of Bulk Synchronous Parallel and MapReduce. Additionally, all approaches were tested on several large graph data sets coming from various domains. 

The initial version of the paper was presented at the ICDM 2012 conference \cite{KajdanowiczICDM2012}.

The following Section \ref{sec:relatedWork} provides a short state-of-the art study on graph problems solutions by means of cloud computing. The main architectures for graph processing including  distributed memory and shared-memory were presented in Section \ref{Sec:ParallelArchitectures}. Two parallel processing models MapReduce and Bulk Synchronous Parallel (BSP) are sketched in the fourth section. Some discussion on their similarities as well as potential improvements is provided in Section \ref{sec:similarities}. Also in this section, an important and experimentally verified MapReduce modification based on map-side join design patterns is proposed for graph processing. Two iterative graph algorithms: single source shortest path computation and collective classification are described more in-depth in Section \ref{sec:Algorithms}. Experimental setup and cloud environment, including data set profiles can be found in Section \ref{sec:ExperimentalSetup}. The results of experiments are presented in Section \ref{sec:experiments}. Discussion on results and solutions of some problems, which arose during research are depicted in Section \ref{sec:Discussion}. The last, tenth section contains general conclusions and further work direction.

\section{Related work}
\label{sec:relatedWork}
Dynamic development of distributed and cloud computing has led to stable solutions for massive data processing. Nowadays, there is an intensified focus on new models useful for specific kind of processing. On top of distributed storage systems many solutions dedicated for particular task are located, for example fast random access data, pipeline processing, graph computations, etc. \cite{hadoop}.

There are several concepts for parallel processing in clusters. Two of them are widely used in offline batch analysis systems and merit special attention: MapReduce and less exoteric Bulk Synchronous Parallel (BSP). The former is especially very popular and applied to many real solutions \cite{hadoop}. 

The general idea behind Bulk Synchronous Parallel (BSP) method was firstly coined and studied in early 90s \cite{valiant}, \cite{Krizanc}. 
Recently, it was adapted by Google to graph processing in clouds in the Pregel system \cite{pregel}. Pregel's idea of using BSP for graph processing in clouds inspired others to create similar systems, some of them are open-source e.g. \cite{giraph}, \cite{hama}.

The overview of large-scale graph engines is presented in \cite{implementations}, which contains graph systems designed to achieve different goals - from offline analytics system to online low-latency systems. 

An empirical comparison of different paradigms for large-scale graph processing is presented in \cite{microsoft}. However, the presented paradigms require a proprietary and / or prototypical platforms, while, in this paper, we focus on approaches which are available on Hadoop, a highly popular, open-source platform, which can be run on a set of commodity hardware.

Pace et al. \cite{pace} provided a theoretical comparison of BSP and MapReduce models. In terms of graph processing, they noticed, that Breadth First Search algorithm (for the shortest path computation) cannot be efficiently implemented by means of the MapReduce model. In this paper, we go forward and focus more on an empirical comparison for the real world data sets, using available implementations as well as evaluation for additional graph problem - collective classification. The general conclusions remain the same: BSP usually appears to be better model for solving graph problems than MapReduce. The results included in this paper provide quantitative analyses supporting that statement.

\section{Parallel Architectures for Graph processing}
\label{Sec:ParallelArchitectures}

Regardless the nature of particular computational problem it can be paralleled and scaled well when the overall solution is balanced in terms of problem solution, algorithm expressing the solution, software that implements the algorithm and hardware. The algorithms, software, and hardware that worked properly for standard parallel applications are not necessarily effective for large-scale graph problems. In general, graph problems have specific properties that make them difficult to fit in existing distributed computational solutions. Among others, following characteristics of graph processing causes challenges in effective parallel processing \cite{Lum2007}:
\begin{itemize}
\item \textbf{Computation driven by data.} Majority of graph algorithms are executed according to structure of a graph, where computation for each next vertex is strictly dependent on results calculated for all antecedents. It means that the the algorithm is rather operating on the graph than executing explicitly stated code. This implies that the structure of whole computation is not known at the beginning of execution and efficient partition is hardly possible.
\item \textbf{Diverse and unbalanced data structures.} Usually graph data is highly unstructured or irregular what do not give much options for parallel processing based on partitioning. Additionally, skewed distribution of vertices degree makes difficult scalability limiting it to unbalanced computational loads.
\item \textbf{High overload for data access in comparison to computation.} Algorithms are often exploring graphs rather then performing complex computations on its structure, e.g. shortest path problem requires only single arithmetic operations in path cost calculation but requires performs many data queries. Runtime can be easy dominated by the wait for memory access, not by computational activities.
\end{itemize}

Due to the fact that commercially available computer appliances have varying capabilities there can be distinguished several processing architectures suitable for distinct hardware. Depending on the amount of available storage and  memory for computation the data might be processed in different manner, reducing or increasing latency. There can be distinguished distributed memory architecture and shared-memory architecture.
 
\subsection{Distributed Memory Architecture}

The most widespread class of parallel computing is distributed memory architecture. This architecture comprises of set of machines (a set of processors and storage/memory) connected by high speed network. It is possible that the environment can be composed of quite common machines and this makes the architecture inexpensive. According to reported results the architecture is effective on many scientific problems but is able to handle only trivially paralleled algorithms.

The general idea behind the distributed memory architecture is to implement the message passing technique. In this method the data is divided into memories of different machines. The distribution of the data is controlled centrally and this means it has to be additionally decided which processor performs which tasks. Usually data from memory is processed by local processor. The distributed memory architecture has a big disadvantage: all tasks to be performed have to be explicitly formulated at he beginning of computation. However it means that the user can almost completely control the computation. Due to the fact, that the data is exchanged between processors by specially designed message passing protocol, the user have to specify it. This makes the architecture very flexible but full control of communication and data partitioning can influence errors. Such problems can be overwhelmed by usage of standards like MPI protopol \cite{MPI:1993}. As long as the architecture enables full customization of implementation smart users can plan such a system realization that achieves high performance. 

One of the best known message passing style in the distributed computing is Bulk-Synchronous one\cite{valiant}. Even thought it is quite mature it has been re-discovered again as powerful implementations\cite{pregel}. In general the processors in this technique switch from working independently on local data to performing collective communication operations. This implies the collective exchange of data unable to be accomplished on demand, but only at the pauses between computational steps (synchronization). It can cause problems with load balancing and actually makes difficult to exploit parallelism in an application. The Bulk Synchronous technique is described in more details in Section \ref{bsp}, where the programming models are considered.

There exists an improvement for message passing technique which is still able to utilize distributed memory - partitioning of global address space. Introducing additional layer of abstraction in implementation that provides operations on remote memory locations with simple manipulation mechanisms facilitates writing programs with complex data access pattern and, therefore, asynchronous communication. An example of partitioned global address space implementation is UPC 
\cite{upc}. 

\subsection{Shared-memory architecture}

There are two possible realizations of shared-memory architecture in computing: hardware based or software based. Both on these approaches are required to provide a support for global memory accessible for all processors. Mentioned in the previous paragraph UPC implementation is actually an example of software providing illusion of globally addressable memory but still working on distinct machines. But the support for a global address space can also be provided in hardware. 

One of the common realizations of shared-memory architecture are symmetric processors which can utilize global memory universally. The architecture assumes, that thanks to proper hardware support any processor can access to addresses in global memory. This feature allows processors retrieve the data directly and relatively in very quick manner. Additionally, solutions of highly unstructured problems, like graph processing, may benefit from that characteristic and achieve higher performance than environments based on distributed memory architecture. 

The shared-memory approach is not ideal and thread synchronization and scheduling reveal another performance challenge. For instance if several threads are trying to access the same region of memory, the system must apply some protocol to ensure correct program execution. Obviously, some threads may be blocked for a period of time. 

Another noticeable problem, that has to be considered while implementing the architecture is the fact, that the best efficiency is obtained when processors are kept occupied with large number of threads. Many graph algorithms can be written with multi-thread fashion, fortunately. However this may imply increases of memory access. 

Finally, the architecture requires processors that are not custom and more expensive than the ones used in distributed-memory architecture. Moreover, the preprocessors  have a much slower clock than mainstream ones. Even the architecture is quite interesting, flexible and more effective it might not be the most attractive for graph processing.

\section{Open Source Parallel Programming Models}
\label{Sec:ParallelModels}

The main purpose do this paper are comparative studies of different practical approaches to parallel graph processing using open source platforms. In particular, the most popular MapReduce (MR), see Section \ref{mapReduceConception} and less common Bulk Synchronous Parallel (BSP), see Section \ref{bsp}, are considered. Additionally, an extended version of MapReduce, namely, map-side join modification of MapReduce (MR2), see Section \ref{sec:mapSideJoin}, together with MR and BSP have been experimentally examined, see Section \ref{sec:ExperimentalSetup} and \ref{sec:experiments}. The overall comparison of these three approaches are described in Table \ref{tab:models}.

{\hfill{}
\begin{table}[htbp]

\setlength{\tabcolsep}{6pt}
\renewcommand{\arraystretch}{1}
\centering
\caption{Profile of various parallel graph processing approaches: MapReduce (MR), its extension based on map-side join design patterns (MR2) and Bulk Synchronous Parallel (BSP)}

{\scriptsize\tt \begin{tabular}{p{3.1cm}p{3.0cm}p{3.0cm}p{2.8cm}}
\addlinespace
\toprule
Feature & MapReduce (MR) & Enchanced MapReduce (MR2) & BSP \\
\midrule
General unit of processing & an iteration composed of 2 Map-Reduce jobs, Figure \ref{fig:MapReduceGraph} & an iteration composed of one Join and one Map-Reduce phases, Figure \ref{fig:MapReduce2Graph} & superstep, Figure \ref{fig:BSPgraph} \\
\\
Data processed & graph structure (neighbourhoods) \& vertex labels & graph structure (neighbourhoods) \& vertex labels & graph structure (neighbourhoods) \& vertex labels\\
\\
Graph vertex allocation & at every iteration, before map & at every iteration, before join & once, at the beginning\\
\\
Work allocation among machines & flexible, repeated at every iteration & flexible, repeated at every iteration & fixed at the beginning \\
\\
Set of vertices processed on single machine & change before each map & change before each map & fixed at the beginning\\
\\
Data about vertex and its neighbours & transferred before each map & transferred before each join & transferred once before entire processing\\
\\
Location of graph data (neighbourhoods \& labels) & HDFS & HDFS & local memory\\
\\
Messages with updated data related to a given vertex (self-update) & transferred before each map & transferred before each join & data stored locally; no message \\

\bottomrule
\end{tabular}}%
\label{tab:models}%
\end{table}}%

\subsection{MapReduce}
\label{mapReduceConception}
\begin{figure}
\centering
\includegraphics[width=0.5\textwidth]{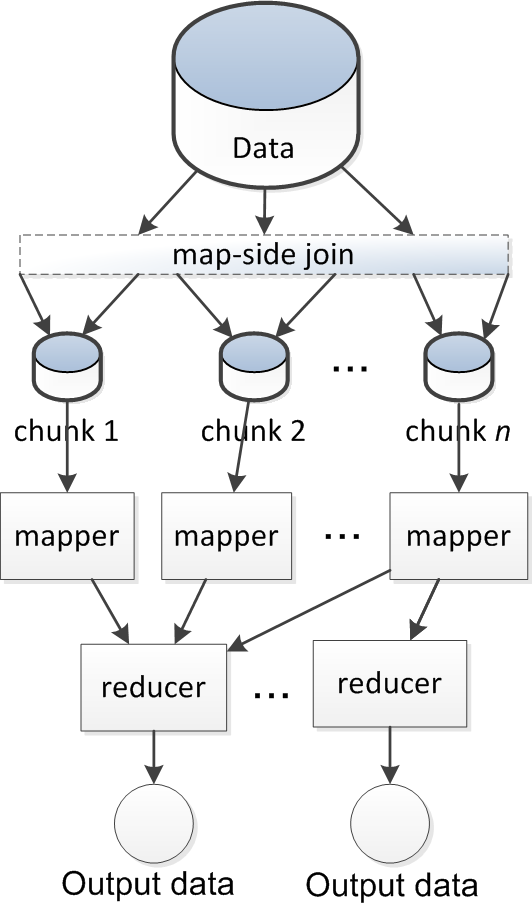}
\caption{Data flow in MapReduce programming model.\label{fig:mapReduce}}
\end{figure}

MapReduce is a parallel programming model especially dedicated for complex and distributed computations, which has been derived from the functional paradigm \cite{hadoop}. In general, MapReduce processing processing is composed of two consecutive stages, which for most problems are repeated iteratively: the map and the reduce phase. The former processes the data on hosts in parallel, whereas the latter aggregates the results. At each iteration independently, the whole data is split into chunks, which, in turn, are used as the input for mappers. Each chunk may be processed by only one mapper. Once the data is processed by mappers, they can emit $<key,value>$ pairs to the reduce phase. Before the Reduce phase the pairs are sorted and collected according to the $key$ values, therefore each reducer gets the list of values related to a given $key$. The consolidated output of the reduce phase is saved into the distributed file system, see Figure \ref{fig:mapReduce}.

The MapReduce model is already an mature concept and although it has not been originally designed to process graphs, a set of design patterns for graph processing have been developed \cite{lin}, \cite{dyer} and \cite{twiddling}. These good practices show how to express iterative algorithms in MapReduce, but they do not overcome the widely recognizable inefficiencies of MapReduce model in networked data processing. 

In case of iterative graph processing algorithms, the graph structure and other static data, which do not change through computation, must be transferred over the network of computational nodes from Mappers to Reducers at each single iteration. It causes a great network overhead and appears to be the greatest pitfall of graphs processing by means of MapReduce. The stateless, two-phased computational nature of MapReduce does not allow vertices to reside on the same host for all iterations. It means that after every map-reduce iteration, the entire data must be written to the global memory in order to be consumed in the next iteration. Since the distributed file system serves as the global memory in the cloud environment, the whole computation is very disk-intensive. Additionally, the map and reduce workers reside on different physical machines and for that reason, the data is constantly transferred over the network, which is the scarce resource in the cloud environment.

\subsection{BSP -- Bulk Synchronous Parallel}
\label{bsp}
To address the problem of MapReduce inefficiency for iterative graph processing, Google has created another a distributed, fault-tolerant system called Pregel \cite{pregel}, which is based on Bulk Synchronous Parallel (BSP) processing model \cite{valiant}. Although Pregel is a proprietary system, it has inspired the creation of several open-source systems, which implement the BSP model like Apache Hama \cite{hama} or Apache Giraph \cite{giraph}.

The computation process in BSP comprises of a series of supersteps (equivalent to MapReduce iterations). In every superstep, a user defined function is executed in parallel on every item from the dataset acting as an agent. Pregel and Pregel-inspired graph analytics systems are vertex-centric: a single agent computation has a graph representation in BSP. It consists of graph vertex identifiers, their current values or states, as well as lists of vertexes' outgoing edges. Before any computation, all graph vertexes are partitioned and loaded into local memories of machines (hosts). They stay there throughout all computation, so that the whole processing is carried out using the local hosts' memories. Graph processing in BSP is organised by means of messages sent between machines hosting individual graph vertexes. At every superstep, each host receives from other hosts the messages related to vertexes preserved by this host and executes a user defined computation function. This function performs local processing on local vertexes and sends messages to some or all vertexes' neighbours in the graph. Once the local computation is finished for a given vertex, processing deactivates itself and the host waits for all other vertices to finish. The barrier of synchronization mechanism allows the next superstep to begin when processing for all vertices is completed in the current superstep. Afterwards, only the vertexes that have received a message are activated.

Since the graph's structure (especially edges and values assigned) does not need to be sent over the network at every superstep (iteration), BSP may be very efficient graph computation model. Only specific messages necessary for the algorithm execution are exchanged between hosts. There is no network overhead related to graph structure passing like in the MapReduce model. Moreover, storing the whole graph structure in local memories of workers allows in-memory computation. The need of disk write-reads as well as objects serialization between iterations is eliminated. However, it is possible only if the graph structure fits in the memories of all workers. Otherwise, the spilling-to-disk techniques must be implemented. Although such techniques are claimed to be implemented in Pregel \cite{pregel}, to our best knowledge, they are not yet supported in its open-source followers. This might be a drawback of choosing BSP model for graph processing, as will be presented later on. 

\begin{figure}
\centering
\includegraphics[width=0.8\textwidth]{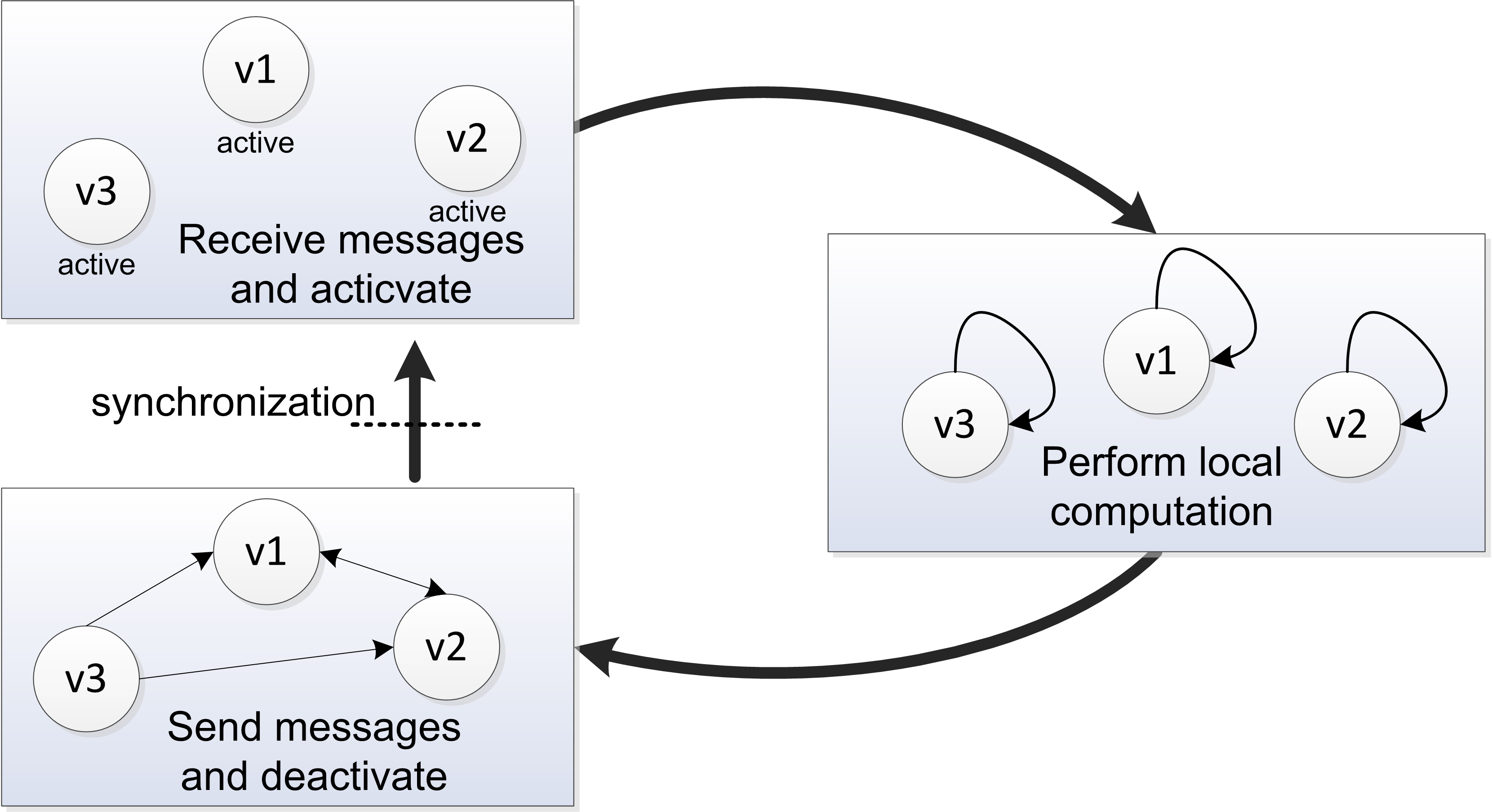}
\caption{Data flow in Bulk Synchronous Parallel programming model.\label{fig:bsp}}
\end{figure}

\section{Similarities Between MapReduce and BSP}
\label{sec:similarities}

Both BSP and MapReduce models differ in many ways but simultaneously they have many common features, which expose them for bottlenecks in graph processing as well as rise possibilities for similar improvements. Some practical enhancements that can be applied for both models are presented in sections \ref{sec:graphPartitioning}, \ref{sec:combiningMessages}. Finally, the new \textit{map-side join} design pattern for MapReduce graph processing is proposed in section \ref{sec:mapSideJoin}. It eliminates passing the graph structure between Map and Reduce phases, bringing two models much closer.

\subsection{Partitioning of the Graph Vertex Set}
\label{sec:graphPartitioning}

By default, in both MapReduce and BSP models, graph vertices are partitioned for the purpose of computation and they are assigned to hosts based on a given, fixed hash-function. This enables the workers-hosts to serve a similar number of vertices. However, it would be beneficial for both models, if graph topologically-close vertices would be processed on the same physical machine. It would increase local computing and decrease network transfer. For that reason, effects of graph partitioning were studied for both the MapReduce model \cite{abadi}, \cite{lin}, \cite{surfer} and for the BSP model \cite{gps}, \cite{mizan}. Commonly, two types of partitioning may be distinguished: (i) range and (ii) cluster partitioning. The former splits vertices using the a-priori knowledge about them, e.g. users from the same geographical region, web sites in the same language, etc. On the other hand, the cluster partitioning strives to extract groups of vertices closely interlinked in the graph. Both of these approaches have two major drawbacks, for which the partitioning aspect was abandon from the research presented in this paper. The range partitioning requires some a-priori knowledge about nodes, which is unavailable, if the source data is anonymized - the case of the data sets used in experiments, see Section \ref{sec:ExperimentalSetup}. The graph vertex clustering, in turn, is a complex and resources-consuming task itself, especially for large graph structures. 

\subsection{Aggregation of Messages}
\label{sec:combiningMessages}
Distribution a workload related to processing of a single vertex is impossible to be optimized in both models. Overall, the entire computation for a given vertex is always executed on a single machine. Since the real world complex network data sets satisfy the power-law distribution for in-degree values, few vertices may require much more processing than the most others. It makes the practical load balancing difficult to achieve both for the MapReduce and SSP model. Additionally, the total computation duration depends on processing time of the mostly loaded computational node and it is valid for both models, even with the perfect balancing and scalability. This problem can be partially addressed by introduction of so called combiners, also to be used both in MapReduce and BSP. Combiners can put together messages destined to any vertex originating from the same machine. They are executed after the map phase in the MapReduce model and after the computation phase, before messages are sent in BSP. Combiners can distribute the workload connected with high-degree vertices and limit the number of messages transferred over the computer network. Unfortunately, there is one significant limitation for this solution - operations on the data must be cumulative and associative, and this condition is not met for all of the graph algorithms. The effect of using combiners for MapReduce is extensively studied in \cite{lin}, while none of equivalent studies is known for graph processing in BSP.

\subsection{Graph Processing using Map-Side Join Design Patterns in MapReduce}
\label{sec:mapSideJoin}

The need of reshuffling the graph structure between Map and Reduce phases is the main disadvantage of graph processing by means of MapReduce. To solve this problem, the Schimmy design pattern was proposed by Lin et al. \cite{lin}. With Schimmy, mappers emit only messages that are being exchanged between vertices. The result of this message processing is merged with the graph structure in the reduce phase and written to the disk in order to be utilized in the following Map-Reduce iteration. To make the merge possible, the key-value pairs representing the graph structure must be partitioned into as many files as the number of reducers and they remain split, similarly as in the MapReduce shuffle phase. Afterwards, every reducer locates and reads (possibly remotely) a file with the range of keys suitable for it.

Instead of a reduce-side join (what actually happens in Schimmy), a better design pattern would be to perform a map-side join. In most graph algorithms, the only time in the MapReduce graph processing when vertices must be aware of the graph structure is in the map phase – when they need to know to which vertices send messages. The reduce phase usually computes new values for vertices and does not require the knowledge about vertex neighbourhoods.

In the map-side join design pattern, the input is partitioned into two groups of files containing either the graph structure or the current state / values of vertices. The input files must be split in the same manner as in case of Schimmy – into the number of files equal to the number of reducers and with the same partitioning functions that routes messages from mappers to reducers. Two types of files are merged in the map phase, i.e. the mapper reads record by record from two files. However, unlike in Schimmy, there is no need to go beyond the standard API – Hadoop offers a special input format that allows the input to consist of more than one file. Mappers emit only the messages exchanged between vertices, likewise Schimmy. Reducers receives vertex messages, perform a computation and emits the result. Output created by reducers will be merged in the next iteration with the graph structure in the map phase. The drawback of this approach is that since mapper will have to read to input files, one of them usually will require a remote read. However, the Schimmy also requires remote reads in the reduce phase, but with the map-side join approach, the graph structure is never written to a disk. It appears to be the greatest advantage over Schimmy. The proposed idea of MapReduce with map-side join, hereinafter referred to as MapReduce 2 (MR2), tends to be noticeably more efficient then the typical MapReduce concept.

The idea of joining both static and dynamic graph data in the map phase has already been proposed in the iMapReduce system \cite{imapreduce}. However, it has not been described in details.

\section{Selected Graph Algorithms Implementing Parallel Processing}
\label{sec:Algorithms}

To evaluate efficiency and other features of different parallel graph processing models, two separate problems were selected. The former is calculation of single source shortest paths (SSSP), see Section \ref{sec:SSSP}. The latter is collective classification algorithm based on relational influence propagation (RIP) used for collective classification of graph vertices, see Section \ref{sec:RIP}. 

\begin{figure}
\centering
\includegraphics[width=0.8\textwidth]{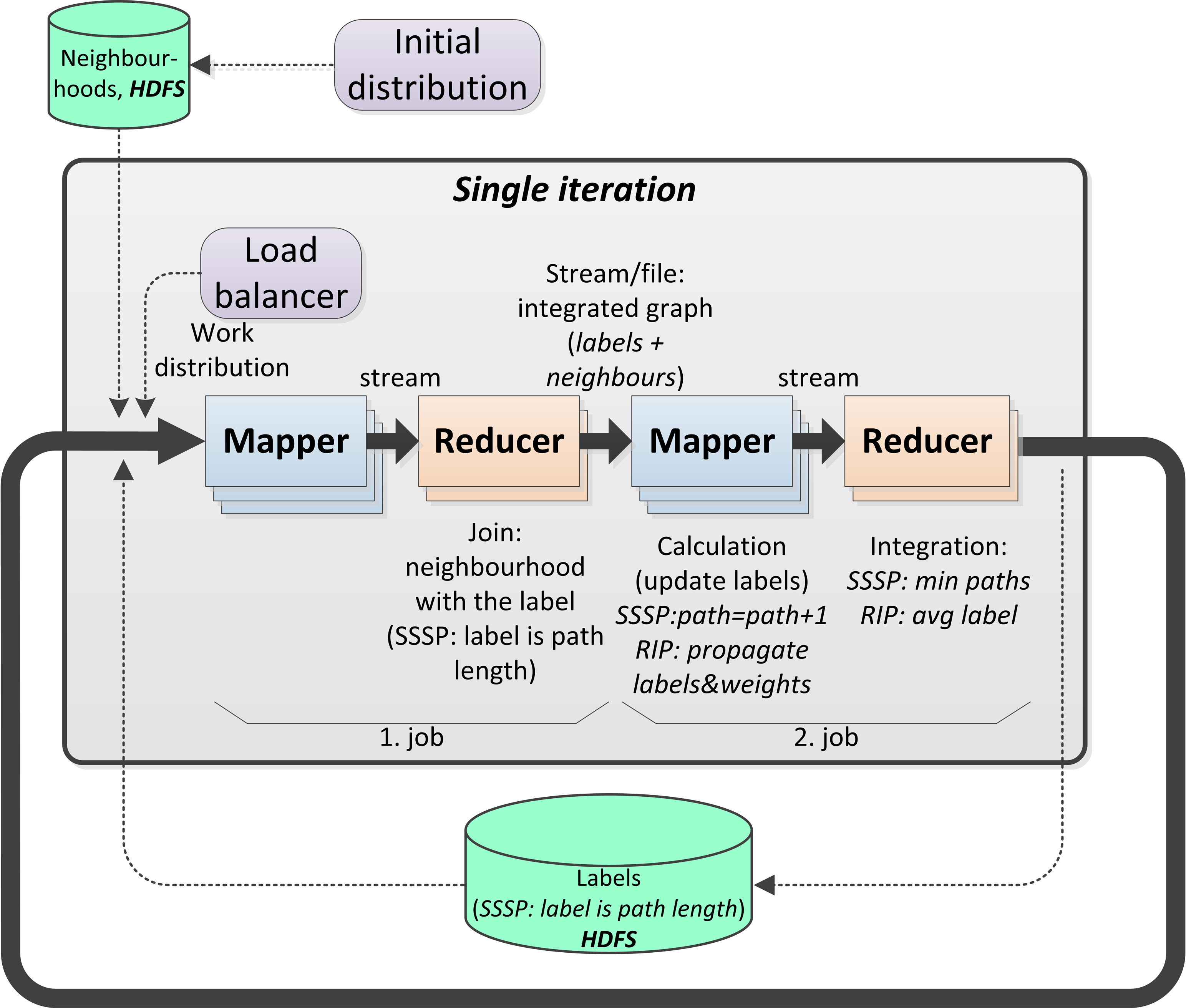}
\caption{MapReduce model for processing of graph problems.\label{fig:MapReduceGraph}}
\end{figure}

\begin{figure}
\centering
\includegraphics[width=0.8\textwidth]{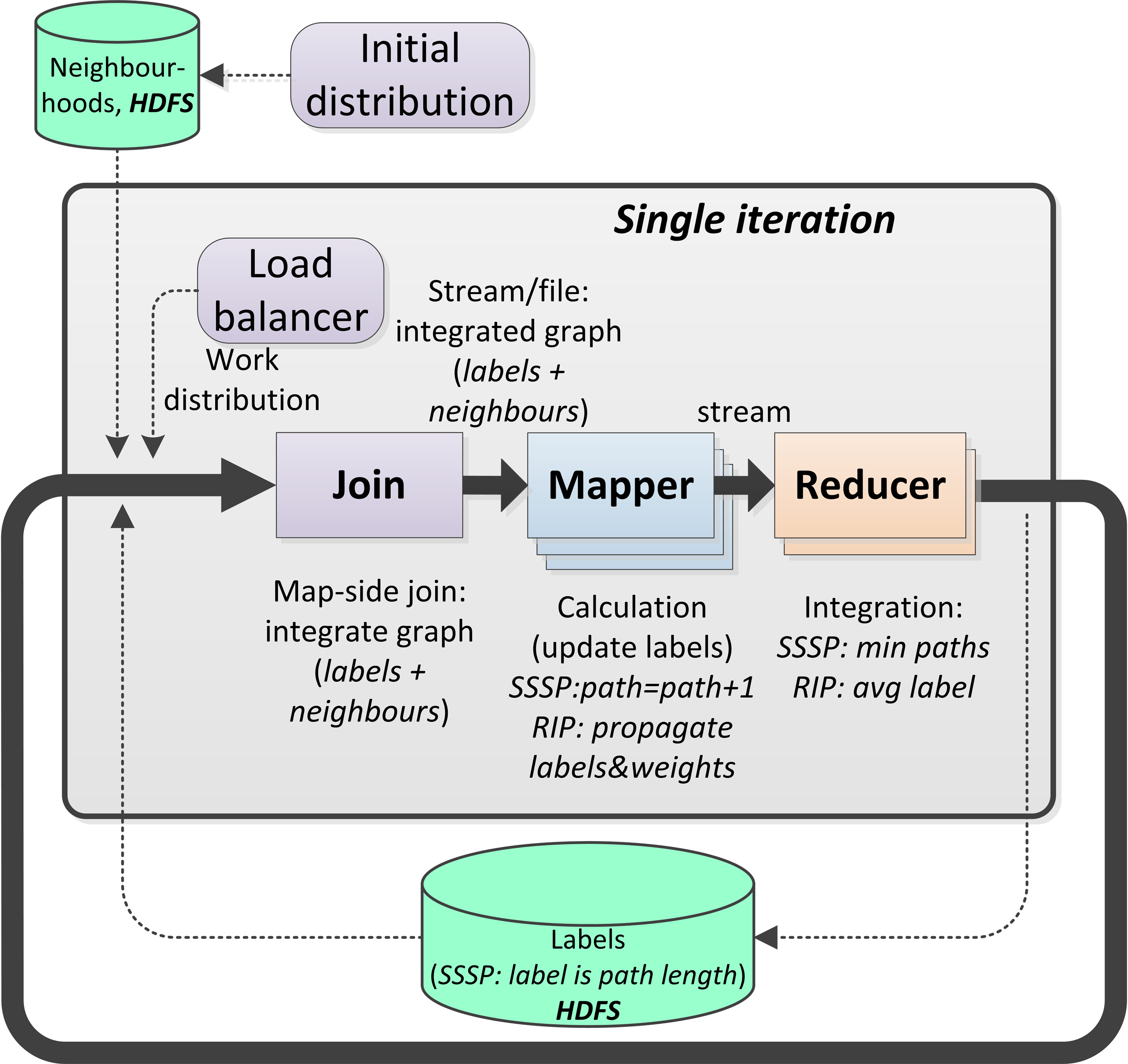}
\caption{MapReduce model with map-side join design patterns (MP2) in graph data processing.\label{fig:MapReduce2Graph}}
\end{figure}

\begin{figure}
\centering
\includegraphics[width=0.8\textwidth]{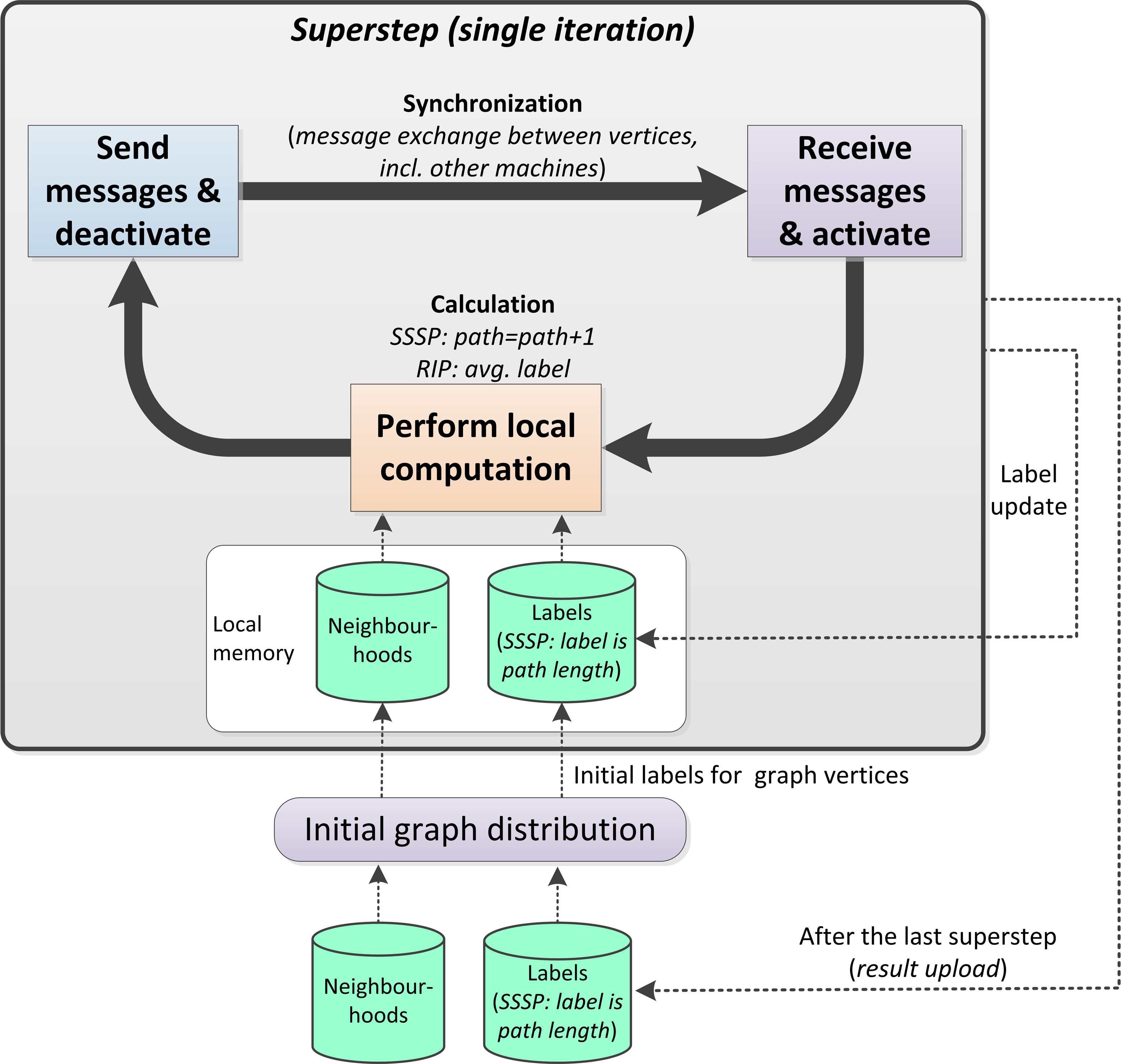}
\caption{Bulk Synchronous Parallel model (BSP) in graph processing.\label{fig:BSPgraph}}
\end{figure}

The general profile of both graph problems are presented in Table \ref{tab:algorithms}. Their three different implementations MapReduce, MapReduce with map-side join extension and Bulk Synchronous Parallel are depicted in Figure \ref{fig:MapReduceGraph}, Figure \ref{fig:MapReduce2Graph}, and Figure \ref{fig:BSPgraph}, respectively.

{\hfill{}
\begin{table}[htbp]

\setlength{\tabcolsep}{6pt}
\renewcommand{\arraystretch}{1}
\centering
\caption{Profile of graph problems used in parallel computation}

{\tt \begin{tabular}{p{6.0cm}p{3.0cm}p{3.0cm}}
\addlinespace
\toprule
Feature & Single Source Shortest Paths (SSSP) & Relational Influence Propagation (RIP) \\
\midrule
Graph type (edges) & unweighted & weighted \\
\\
Graph structure processed independently in parallel within a single iteration & individual vertices with their closest neighbours & individual vertices with their closest neighbours \\
\\
Vertex activation (label update and further broadcasting) & only if the message from the neighbour contains smaller path length & for all vertices \\

\bottomrule
\end{tabular}}%
\label{tab:algorithms}%
\end{table}}%

\subsection{Single source shortest path}
\label{sec:SSSP}

A typical graph problem that was widely studied in graph analysis is the single source shortest paths (SSSP) calculation. For unweighed graphs, this problem can be interpreted as computation of the minimal number of edges, which form a path from an arbitrary chosen vertex to every other vertices in a graph. Implementations of the algorithm have been studied and proposed both for MapReduce \cite{dyer} and for BSP \cite{pregel}.

At the beginning, all vertices are initiated with the maximum possible value and the source vertex with value 0. In the first iteration, the source vertex sends an updated shortest path value equal one to its immediate neighbours. In the next iteration, these neighbours may propagate the incremented by 1 path lengths to their neighbours, and so on. Every vertex (neighbour) may receive a message with the shortest path smaller than its currently stored value. In such case, the vertex becomes active and broadcasts the updated shortest path (increased by 1) to all its neighbours. However, if the received path length is greater than the current value for a given vertex, it does not send any message. In this way at every iteration (MapReduce job / BSP superstep) a frontier of vertices with the newly discovered shortest paths is extended by one hop. The entire algorithm can be simply translated from the MapReduce to BSP model and vice versa. The MapReduce processing is expressed in two functions and the Map functions emit the vertex itself to preserve the graph structure.

\subsection{Collective Classification of Graph Vertices based on Relational Influence Propagation}
\label{sec:RIP}

Overall, the term \textit{collective classification} refers to the specific classification task of a set of interlinked objects. The common representation used in such problems are graphs, where each object is represented by a vertex and relationships between objects are graph edges. In collective classification, each vertex can be classified according to the generalized knowledge derived from correlations between the labels and attributes of vertices or correlations between the labels and observed attributes supplemented by labels of neighbouring vertices. Unlike standard classification, collective approach utilizes information about relationships between vertices. For example, it is very likely that a given web page $x$ is related to music (label \textit{sport}), if page $x$ is linked by many other web pages about sport.

Collective classification can be accomplished by two distinct inference approaches: within-network classification or across-network classification. Whereas the former tries to recover partially unknown labels in the network based on information of known ones, the latter applies the recovered label distribution from one network to another.

The within-network collective classification algorithm based on \textit{Relational Influence Propagation} (RIP) has been examined in this paper. The general idea of this approach is based on the iterative propagation of the known labels assigned to some known vertexes to the non-labelled vertices in the same network. The method was originally derived from the enhanced hypertext categorization \cite{Cha98}.

The RIP algorithm for MapReduce and BSP have already been proposed in \cite{Indyk:2012:MTAP} and \cite{bsploopy}, respectively. The pseudo-codes of implementations for MapRaduce and BSP are presented in Algorithm 1 and 2, respectively, in order to demonstrate the differences between these two programming models. The implementation in the MapReduce paradigm consists of two separate functions: \textit{map}() and \textit{reduce}(). Each \textit{map}() reads and processes vertex by vertex. For every vertex with the known label, it propagates its label to its all immediate neighbours (lines 2-3 in Algorithm 1). Since MapReduce is the stateless model, it must also send its own representation to itself (line 4). Messages transmitted by the map functions are consumed by the reduce functions, which are usually executed on different machines. When the reduce function receives all the messages, it must distinguish messages from two types: a message sent by itself, which contains the vertex adjacency list (lines 4-5) and messages sent by its direct neighbours having their labels already assigned. Next, it aggregates all received external labels along weights of graph edges linking the vertex to its neighbours (lines 7-8). Finally, the updated likelihood expressed by means of the weighted mean of neighbourhood likelihood is computed for a given vertex (line 11). As a result, the vertex obtains its own new label. At the end, the vertex propagates its own representation containing its updated label. The output will serve as the input for the map function in the next iteration (MapReduce job).

Unlike MapReduce, in the BSP implementation, the whole processing for a vertex is expressed with only one function \textit{compute()}, which is executed on a single machine, see Algorithm 2. Function \textit{compute()} is triggered once in each superstep (iteration) separately for every vertex that has received a message sent in the previous superstep. The logic of the RIP algorithm is expressed similarly as in MapReduce: vertices receive and aggregate messages from their neighbours (line 2-5), calculate the new likelihood (line 6) and send their updated likelihood to all their neighbours. The main difference between the MapReduce model is that the vertices do not have to propagate themselves. This is guaranteed by the stateful nature of processing in BSP -- all the computation for one vertex is always executed on the same machine and the vertex state including information about its neighbours (in particular weights of edges) can be stored in-memory or another local storage.

\begin{spacing}{1}
\begin{algorithm}
\caption{MapReduce approach to Relational Influence Propagation}
\label{alg:collectiveMapReduce}
{\tt\begin{algorithmic}[1]
\Function{map}{vertexId n, vertex V}
	\For{outgoingEdge E $\in$ V.adjacencyList}
		\State emit(E.neighborId, $<$V.label, E.edgeWeight$>$)
	\EndFor
	\State emit(vertexId n, vertex V)
\EndFunction
\item[]
\end{algorithmic}
\begin{algorithmic}[1]
	\Function{reduce}{vertexId n, messages[$m_1 , m_2 , ... $]}
		 	\State originalVertex V = null
		 	\For{m $\in$ messages}
				\If{m.isVertex()}
					\State V $\leftarrow$ m
				\Else
					\State sumLabels += m.label*m.weight
					\State sumWeights += m.weight
				\EndIf
		 	\EndFor
		 	\State V.label $\leftarrow$ sumLabels/sumWeights
		 	\State emit(vertexId n, Vertex V)
	\EndFunction
\end{algorithmic}}
\end{algorithm}
\end{spacing}

\begin{spacing}{1}
\begin{algorithm}
\caption{Bulk Synchronous Parallel approach to Relational Influence Propagation}
\label{alg:collectiveBsp}
	{\tt\begin{algorithmic}[1]
	\Function{compute}{vertex V, messages[$m_1,m_2, ... $]}
		\For{m in messages}
			\State sumLabels += m.label*m.weight
			\State sumWeights += m.weight
		\EndFor
		\State V.label $\leftarrow$ sumLabels/sumWeights
		\For{E $\in$ V.adjacencyList}
			\State sendMsg(E.neighborId, [V.label, E.edgeWeight])
		\EndFor
		\State voteToHalt()
	\EndFunction
	\end{algorithmic}}
\end{algorithm}
\end{spacing}

\section{Experimental Environment}
\label{sec:ExperimentalSetup}

The main goal of experiments was to validate and compare open source parallel programming models: (1) MapReduce (MR), (2) MapReduce with map-side join (MR2) and (3) Bulk Synchronous Parallel (BSP), see Section \ref{Sec:ParallelModels}. These three approaches: MR, MR2 and BSP were evaluated in terms of computational complexity for distinct settings of distributed environment. The efficiency measures were recorded for clusters with various number of computational nodes (from 10 to 85 machines). The analysis was performed for distinct real world datasets and for two graph analysis problems: SSSP and RIP, see Section \ref{sec:Algorithms}. Since all implementations required the same number of algorithm iterations (equal to 10), the mean execution time of a single iteration was used as the evaluation measure in order to compare all approaches.

\subsection{Cloud Setup}
The experiments were carried out using cluster environment provided by the Wroclaw Networking and Supercomputing Center. The distributed system consisted of 85 identical machines with 4 CPUs, 7500MB RAM and hundreds GB of storage each. Computers were connected through 1Gb/s Ethernet. The experiments were deployed on Hadoop (version 0.20.205.0) and Apache Giraph (version 0.2). Apache Giraph \cite{giraph} is an immature, open-source implementation of the BSP model for graph processing in the cloud environment. All programs for Giraph were written as a Hadoop jobs because Giraph launches BSP workers within mappers and then uses Hadoop RPC to enable workers communicate with each other.

\subsection{Datasets}

To compare the MapReduce and BSP approaches four separate datasets were used: \textit{tele}, \textit{tele\_small}, \textit{youtube} and \textit{twitter}. The \textit{tele} dataset is a network of telecommunication clients built over 3 months history of phone calls from one of leading European telecommunication company. The raw dataset used to construct the network consisted of about 500 000 000 phone calls and more than 16 million unique users. Extracted network was built using activities of clients performing calls in two most popular from 38 tariffs. Another dataset, \textit{tele\_small}, was composed based on the next two most common tariffs. In both datasets users were marked with empirical probability of tariffs they used, namely, the sum of outcoming phone calls durations in particular tariff was divided by summarized duration of all outcoming calls. 


The YouTube dataset \cite{Cheng:2008} was crawled using YouTubeAPI in 2008. A subset of all attributes was used to create a weighted graph structure and labels:  \textit{video\_id, age, category, related\_IDs}. Using \textit{related\_IDs} the connection strengths were calculated as a fraction of the number of related videos adjacent videos, i.e. if there were 20 related videos each of them was linked by an edge with the weight of $0.05$. 

The Twitter dataset was the largest dataset utilized in the experiments. It contained a list of two users' identifiers, implying a link from the first to the second user (the first user follows the second one). As there were not available any labels for a user, to enable RIP validation a binary class was assigned randomly to each of network vertex using the uniform distribution. Some other details about the datasets are presented in Table \ref{tab:datasets}.

{\hfill{}
\begin{table}[htbp]
\setlength{\tabcolsep}{6pt}
\renewcommand{\arraystretch}{1}
\centering
\caption{Description of datasets used in experiments}

{\tt \begin{tabular}{p{3cm}p{2.1cm}p{2.1cm}p{2.1cm}p{2.1cm}}
\addlinespace
\toprule
 & tele\_small & tele & youtube & twitter \\
\midrule
Domain & telecom & telecom & multim. & microblog \\
No. of nodes & 5,098,639 & 13,914,680 & 16,416,516 & 43,718,466 \\
No. of edges & 21,285,803 & 67,184,654 & 66,068,329 & 688,352,467 \\
Avg deg. & 4.17 & 4.83 & 4.02 & 15.75 \\
Max indegree & 40,126 & 294,690 & 4,104 & 1,228,086 \\
Direction & directed & directed & directed & directed \\
Classes & 2 & 2 & 15 & - \\
\bottomrule
\end{tabular}}%
\label{tab:datasets}%
\end{table}}%

\section{Results of Experiments}
\label{sec:experiments}

\subsection{Time of Computation for Various Graph Problems}

\begin{figure}
\centering
\includegraphics[width=0.7\textwidth]{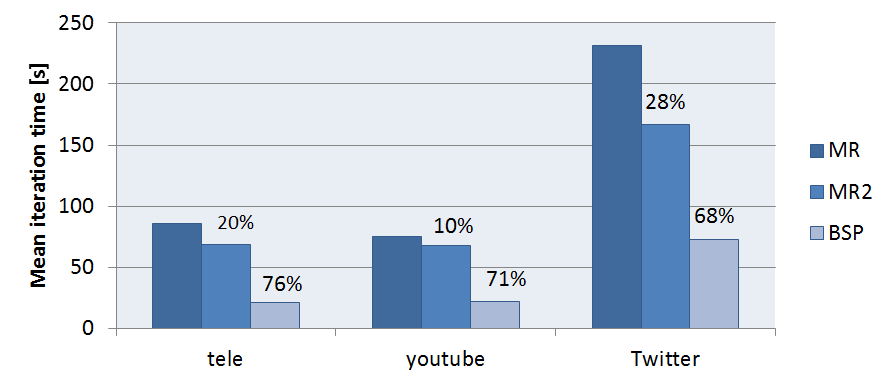}
\caption{Performance of Relational Influence Propagation problem on 60 machines.\label{fig:algCompCC}}
\end{figure}

\begin{figure}
\centering
\includegraphics[width=0.7\textwidth]{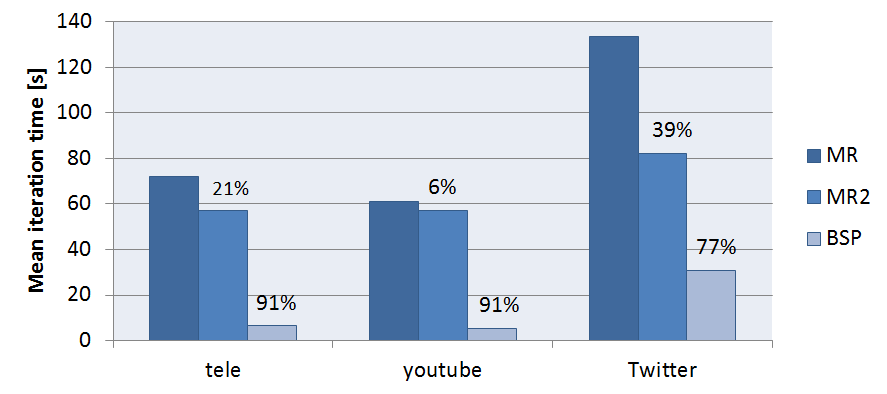}
\caption{Performance of Single Source Shortest Paths on 60 machines.\label{fig:algCompSSSP}}
\end{figure}

First of all, MR, MR2 and BSP approaches were examined against the type of graph processing problems using the fixed number of computational nodes (60 machines). The Relational Influence Propagation (RIP) as well as Single Source Shortest Paths (SSSP) problems were executed in order to test the execution time. The results for three parallel approaches separately for the RIP and SSSP problems are presented in Figures \ref{fig:algCompCC} and \ref{fig:algCompSSSP}, respectively. In general BSP outperformed the regular MapReduce (MP) by 70\%-90\% (3-10 times better) and the improved MapReduce (WR2) by 50\%-90\% (2-10 times better) and more for the SSSP problem than for RIP. Note also that the efficiency gain obtained by MR2 is slightly higher for the SSSP problem than for RIP.

\subsection{Scalability in Relation to the Graph Size}

\begin{figure}
\centering
\includegraphics[width=0.6\textwidth]{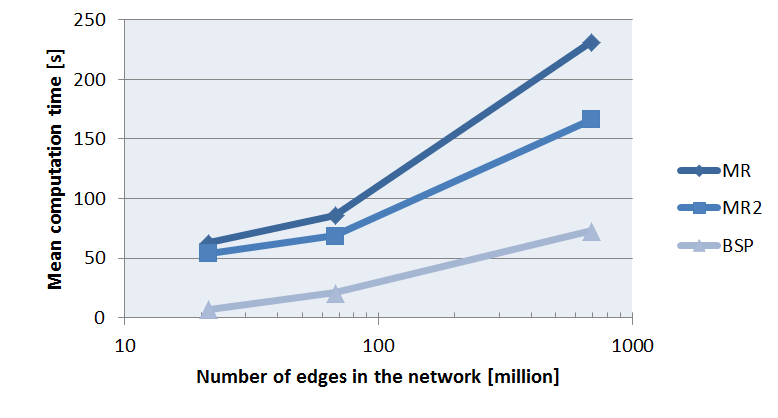}
\caption{Scalability with respect to the size of graph for Relational Influence Propagation algorithm with the fixed number of machines (60).\label{fig:scalabilityCC}}
\end{figure}

\begin{figure}
\centering
\includegraphics[width=0.6\textwidth]{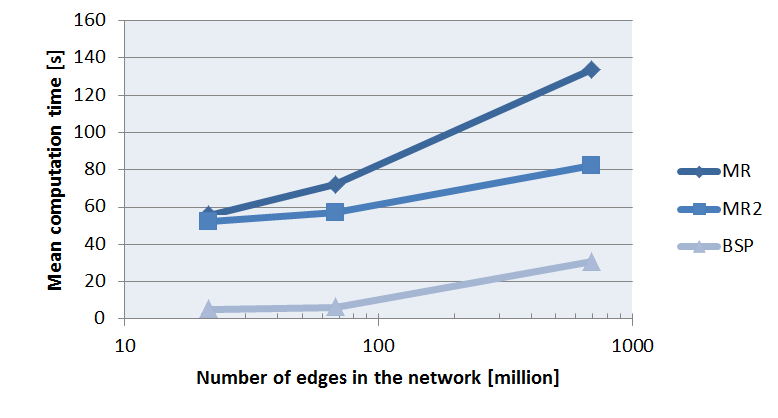}
\caption{Scalability with respect to the size of graph for Single Source Shortest Path algorithm with the fixed number of machines (60).\label{fig:scalabilitySSSP}}
\end{figure}

The next research focused on the question, how the processing time of MR, MR2 and BSP solutions depends on the sizes of graphs for the fixed configuration (60 machines). Again, the experiment was performed on three datasets with significantly different number of nodes (\textit{tele\_small}, \textit{tele} and \textit{twitter}). According to Figure \ref{fig:scalabilityCC} and Figure \ref{fig:scalabilitySSSP} all three approaches shown their nearly linear scalability for both graph problems. Once again, MR2 and BSP outperformed MR implementation, especially for the largest graph, even though the gain is quite stable. The similar characteristics was observed in relation to the increasing number of edges.

\subsection{Horizontal Scalability in Relation to the Number of Parallel Machines}
\label{sec:subSecHorizontal}

\begin{figure*}

\begin{tabular}{cc}

\includegraphics[width=0.45\textwidth]{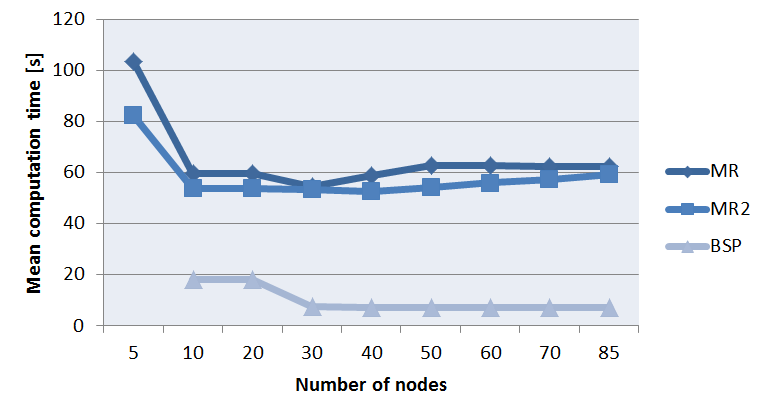}
\label{fig:scalabilityTeleSmallCC}
&
\includegraphics[width=0.45\textwidth]{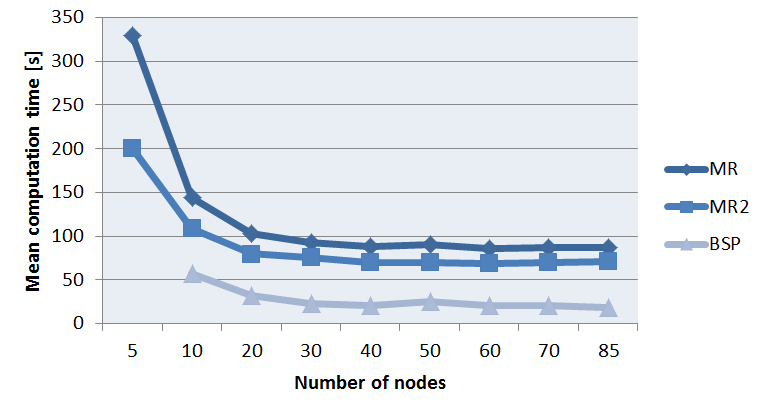}
\label{fig:scalabilityTeleCC}
\\
(a) & (b)
\\
\includegraphics[width=0.45\textwidth]{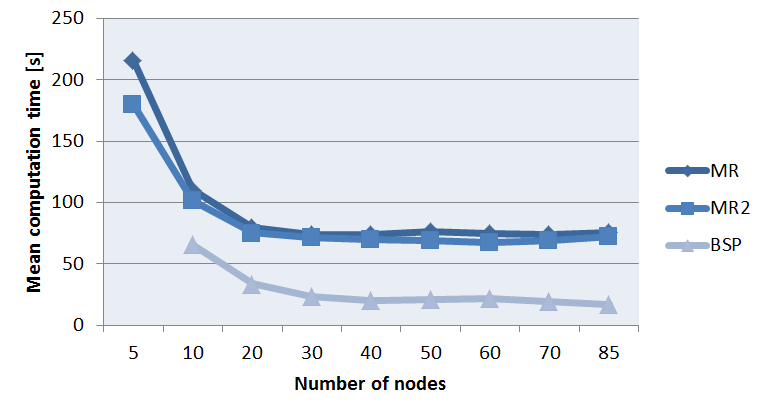}
\label{fig:scalabilityYtCC}
&
\includegraphics[width=0.45\textwidth]{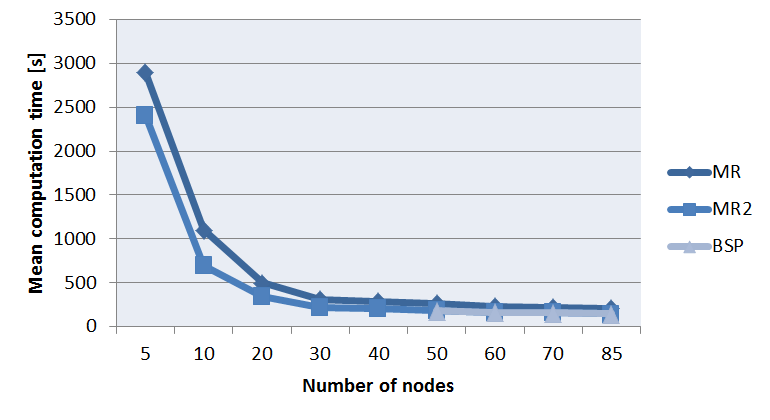}
\label{fig:scalabilityTwitterCC}
\\
(c) & (d)
\\
\includegraphics[width=0.45\textwidth]{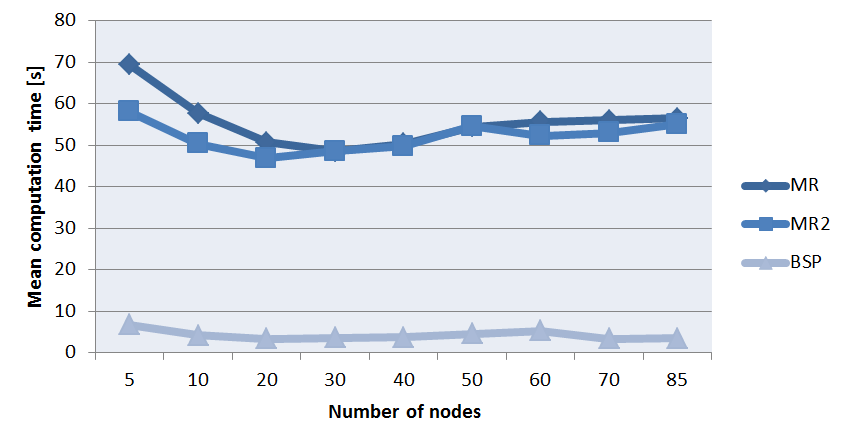}
\label{fig:scalabilityTeleSmallSSSP}
&
\includegraphics[width=0.45\textwidth]{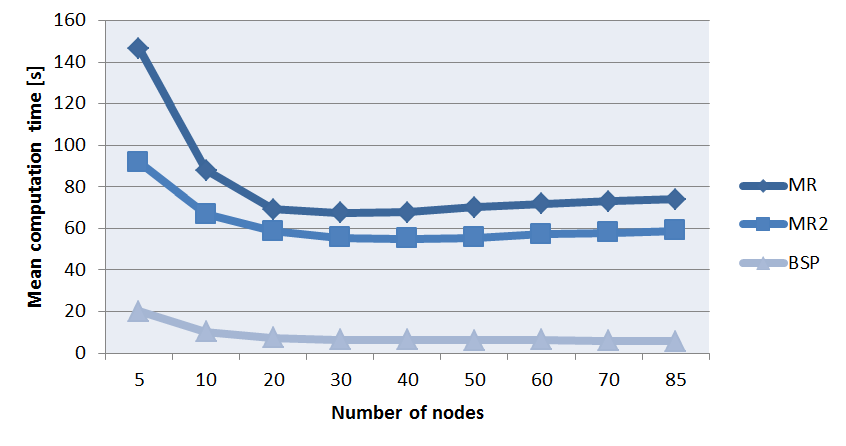}
\label{fig:scalabilityTeleSSSP}
\\
(e) & (f)
\\
\includegraphics[width=0.45\textwidth]{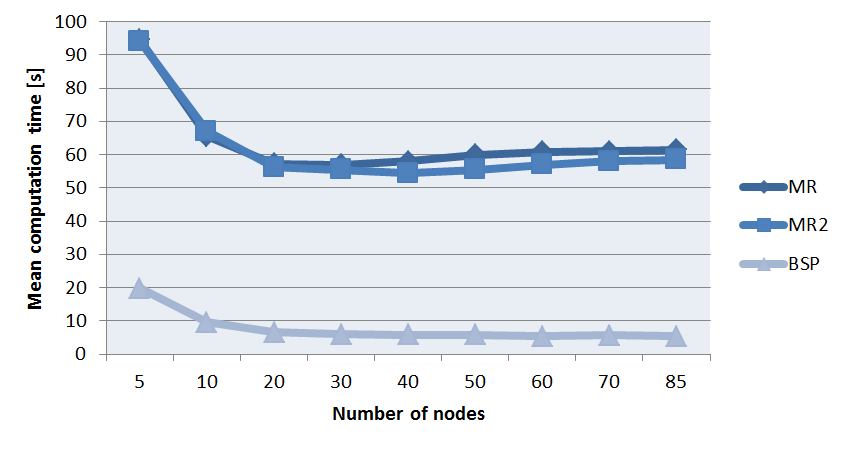}
\label{fig:scalabilityYtSSSP}
&
\includegraphics[width=0.45\textwidth]{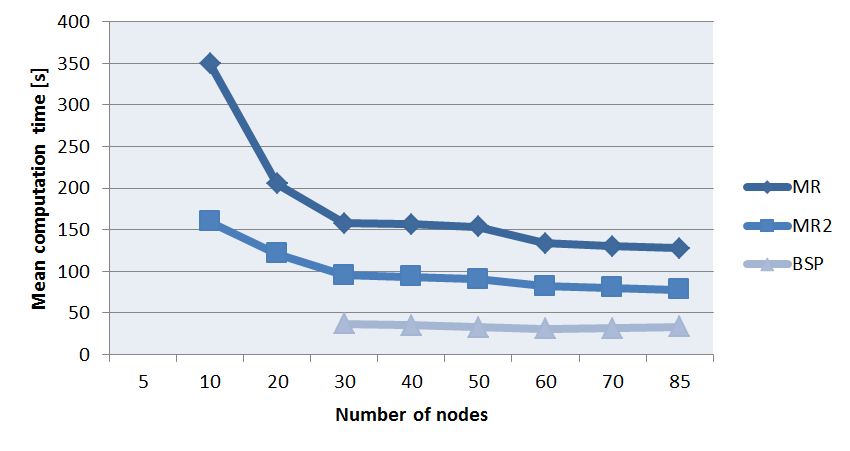}
\label{fig:scalabilityTwitterSSSP}
\\
(g) & (h) 
\end{tabular}%

\caption{Scalability with respect to the number of hosts (nodes) in the computer cluster for the Relational Influence Propagation (RIP) algorithm applied to \textit{tele\_small} (a), \textit{tele} (b), \textit{youtube} (c), \textit{twitter} (d) datasets as well as for the Single Source Shortest Path (SSSP) algorithm executed on \textit{tele\_small} (e), \textit{tele} (f), \textit{youtube} (g), \textit{twitter} (h) datasets.\label{fig:scalabilityAllProb}}
\end{figure*}

The influence of the computer cluster (cloud) size on efficiency (processing time) was also examined. Three parallel models (MR, MR2 and BSP) were compared against each other for different number of machines starting from 5 to 85. The results are depicted in Figure \ref{fig:scalabilityAllProb} (a-h). 

For the smaller datasets like \textit{tele\_small}, \textit{tele}, \textit{youtube} all three solutions, MR, MR2 and BSP, converge to limit of the best mean computation time of iteration with about 30 computational nodes (hosts) in the cloud. Adding more machines eliminates the benefit of additional computing power due to the overhead related to network communication. The mean execution time of a single iteration can be decreased from 2 to 4 times by increasing the number of computational nodes from 5 to 30. This benefit is relatively higher for the RIP problem than for SSSP and in most cases it is higher for BSP than for MapReduce approaches. The 17 times increase of the number of machines (from 5 to 85) speeds up the processing 14-times (2,900 sec. to 205 sec.) for MR, the RIP problem and \textit{twitter} (Figure \ref{fig:scalabilityAllProb}d) and even 16-times for MR2 (2,400 sec. to 146 sec.). Due to the in-memory calculation limit of the BSP approach, it was able to execute the \textit{twitter} dataset on environment starting only from 50 machines (Figure \ref{fig:scalabilityAllProb}d). Adding more machines improves the performance only slightly. 

On the other hand, the mean time of one iteration depends only to a moderate degree on the number of computational nodes; it refers to computing for the SSSP problem (Figure \ref{fig:scalabilityAllProb}e-h). The biggest observed improvement for MR and MR2 was only 2.7 and 4.6 times faster, respectively. 

Similarly to the RIP algorithm for the \textit{twitter} dataset, the BSP implementation of SSSP was able to be executed not until 30 or more machines in the cluster (Figure \ref{fig:scalabilityAllProb}h). Once more it was observed, that in case of BSP approach, increasing the number of computational nodes reveals only slight improvement of a processing time.

The worst scalability can be observed for the \textit{tele\_small} dataset, less than 2 times for MP and MR2 while increasing the number of hosts from 5 to 10, and about 3 times for BSP but not until 20-30 nodes (Figure \ref{fig:scalabilityAllProb}a,e).

For some cases, the extension of the computer cluster over 30 machines even decreases efficiency -- average processing time  slightly rises (Figure \ref{fig:scalabilityAllProb}a,e,f,g), but it is valid only for MapReduce solutions (MR and MR2). In these cases, efficiency stabilizes at the certain level for BSP .

\subsection{Scalability in Relation to the Number of Algorithm Iterations}
\label{sec:subSecIterations}

\begin{figure}
\centering
\includegraphics[width=0.6\textwidth]{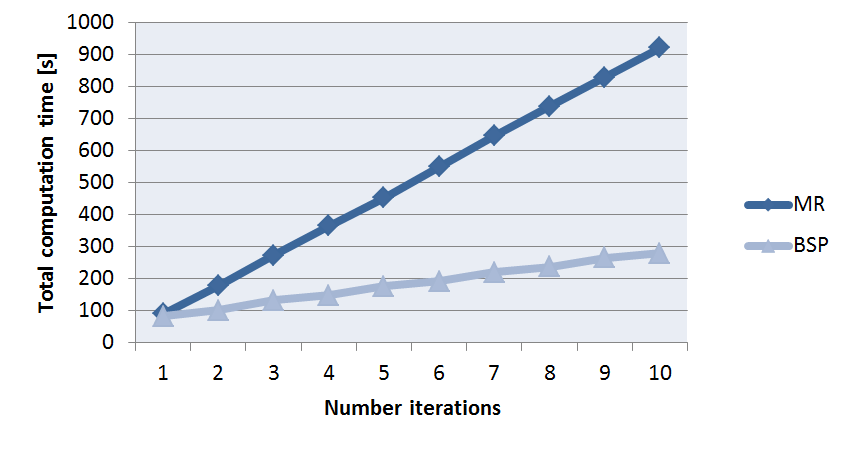}
\caption{Total processing time with respect to number of iterations in Relational Influence Propagation algorithm run on 40 machines.\label{fig:scalabilityIterCC}}
\end{figure}

\begin{figure}
\centering
\includegraphics[width=0.6\textwidth]{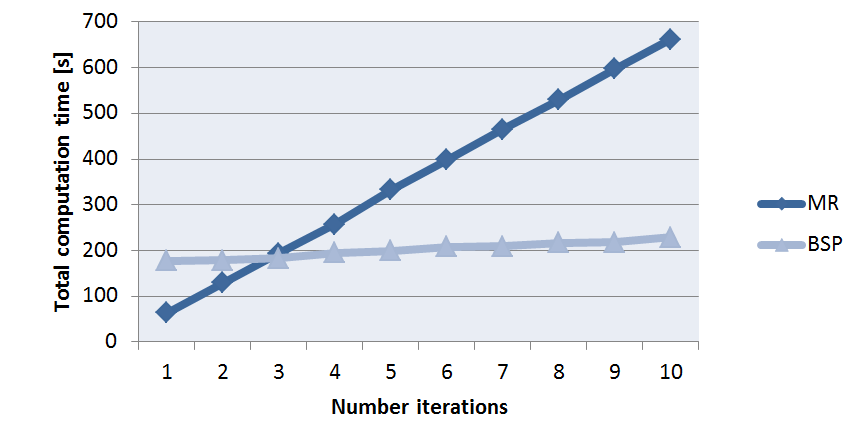}
\caption{Total processing time with respect to number of iterations in Single Source Shortest Path algorithm exectuted on 40 machines.\label{fig:scalabilityIterSSSP}}
\end{figure}

The total processing time for RIP and SSSP algorithms was shown in Figures \ref{fig:scalabilityIterCC} and \ref{fig:scalabilityIterSSSP}. The values presented additionally aggregate the initiation time attributed to each iteration for BSP and MapReduce models. In both approaches, regardless of the calculated problem, the processing time increases linearly with the number of iterations. The coefficient of determination ($R^2$) between real values and their linear equivalence is for MR at the level of 0.9998-0.9999 and for BSP: $R^2$ = 0.9747-0.9976. A bit lower correlation between real values and linear approximation for BSP compared to MR was caused by w bit longer duration of the first iteration (superstep) of BSP, which contained the graph data distribution among parallel machines.

Additionally, MapReduce approach iterations are significantly slower and their total processing time rises more rapidly. 

It can be also noticed, that the time required to complete the first iteration is 3 times longer for BSP than for MR implementation for the SSSP problem. It is caused by the initialization procedure required for loading the graph structure into the distributed memory before the first superstep of BSP. It is valid for the SSSP problem, however, for RIP, the first iteration lasts the same time for both approaches. It means that the time consumed for loading structure of the graph may vary on many factors and it is strongly implementation-dependent. 

\section{Discusion}
\label{sec:Discussion}

In general, the results presented in Section \ref{sec:experiments} revealed that the MapReduce approach (MR) is outperformed by Bulk Synchronous Parallel (BSP) as well as by application of the map-side join design pattern (MR2). Nevertheless, the efficiency gain differs for distinct computational problems. It is exposed that usage of MR2 or BSP instead of raw MR provides greater performance improvement for Single Source Shortest Paths (SSSP) calculation than for Relational Influence Propagation (RIP) computation. This can be explained by the time required for communication between computational nodes (hosts). In RIP, the communication was very dense as 70\% of all graph vertices used as a source for propagation, were propagating their labels to almost all other vertices. In the SSSP problem, in turn, the number of propagating vertices was changeable and depended on the iteration, but in general, was much smaller than in RIP. 

The general concept of the BSP model facilitates in-memory computation for the entire graph. This is the main reason, why the BSP implementation outperforms MapReduce for all considered graph-based problems. In case of MapReduce, the messages exchanged between vertices as well as intermediate data transferred in-between iterations must be written to the disk. However, in-memory processing requires that the whole graph structure and all intermediate data produced during processing (e.g. messages exchanged between vertices) must fit in the memory. Otherwise, a spilling mechanism should be provided in order to manage the data in the memory and external storage. Unfortunately, to the best of our knowledge, recently none of the open-source BSP systems provides any equivalent mechanism.  Hence, for very large networks, MapReduce remains the only alternative. As it was shown in Section \ref{sec:subSecHorizontal}, the risk of not fitting in memory is higher for more communication intensive problems like Relational Influence Propagation. 

Overall, the in-memory processing allows graph's vertices to remain in the same physical location throughout the computation, see Figure \ref{fig:BSPgraph}. On the other hand, additional overhead is required at the beginning of computation in order to partition the network and load it into memory of workers. The overhead may be compensated after few iterations (as was discussed in Section \ref{sec:subSecIterations}, see Figure \ref{fig:scalabilityIterSSSP}). Notwithstanding, the BSP model may be a very good solution for non-iterative graph algorithms.

The necessity of data exchange and access to distributed file systems (HDFS) grows with the higher number of parallel machines. It especially refers MapReduce approaches, both original (MR) and improved (MR2), since the graph vertex may be re-allocated at every iteration of MapReduce-based algorithms. As a result, adding new machines does not need to decrease average processing of a single algorithm iteration. For the datasets and graph problems analysed in experiments, the upper reasonable limit of computational nodes is about 20-30, see Section \ref{sec:experiments} and Figure \ref{fig:scalabilityAllProb}.

\section{Conclusion and Further Work}

Three main approaches to parallel processing of graph data, namely: (i) MapReduce (MR), along with (ii) its extension based on map-side joins (MR2) as well as (iii) the Bulk Synchronous Parallel (BSP) were examined in the paper. Two graph problems, that can be solved by means of iterative algorithms were implemented and tested separately for the above three approaches: calculation of lengths of single source shortest paths (SSSP) as well as relational influence propagation (RIP) used for collective classification of graph nodes. 

The experimental studies on four large graph datasets with different profiles revealed that Bulk Synchronous Parallel approach outperforms other solutions (MR and MR2) for all datasets and all tested iterative graph problems. The BSP model, despite of its relatively young implementation, worked even up to one order of magnitude faster than the MapReduce-based approaches. Superiority of BSP was greater for greater for telecom data rather than for twitter data (compare Figure \ref{fig:scalabilityAllProb}a,e with Figure \ref{fig:scalabilityAllProb}d,h).

Simultaneously, it was discovered that the map-side join design pattern (MR2) may improve the original MapReduce performance up to two times. It is caused by replacement of one  Map-Reduce job by single join at every iteration (compare, Figure \ref{fig:MapReduceGraph} and \ref{fig:MapReduce2Graph}).

Parallel processing of graph data has also some limitations. The gain in increasing the number of parallel hosts is visible only up to certain quantity. Based on experiments from Section \ref{sec:experiments}, see Figure \ref{fig:scalabilityAllProb}, we can state that about 20-30 hosts is the upper limit. The increase above this threshold does not result in faster processing. it is visible especially for MapRedure solutions (MR and MR2). This phenomena was caused by necessity of the more extensive data exchange in case of larger number of parallel machines.

Nevertheless, MapReduce can still remain the only alternative for parallel processing of graph algorithms on huge datasets. It results from the main BSP limitation: the very high memory requirements -- all the data used for local processing in BSP must fit in the local memory.

\section*{Acknowledgment}
This work was partially supported by The Polish National Center of Science the research project 2010-13 and 2011-14.



\bibliographystyle{elsarticle-num}
\bibliography{fgcs}







\end{document}